\documentclass[mathleft
]{an}
\usepackage{graphicx}
\usepackage{times}
\begin{document}

\Pagespan{789}{}
\Yearpublication{2010}%
\Yearsubmission{2010}%
\Month{}%
\Volume{}%
\Issue{}%

\title {Eclipsing binaries with pulsating components: CoRoT 102918586}

\author{C. Maceroni\inst{1}\fnmsep\thanks{Corresponding author:
  \email{maceroni@oa-roma.astro.it}\newline}, D. Cardini\inst{2} , C. Damiani\inst{1}, D. Gandolfi \inst{3,4}, 
  J. Debosscher \inst{5}, A. Hatzes \inst{4}, E.W. Guenther \inst{4} \and C. Aerts \inst{5}
}

\titlerunning{Eclipsing binaries with pulsating components: CoRoT 102918586}
\authorrunning{C. Maceroni et al.}
\institute{
INAF--Osservatorio astronomico di Roma, via Frascati 33,I-00040 Monteporzio C., Italy
\and 
INAF, IASF--Roma,Via del Fosso del Cavaliere 100, 00133 Roma, Italy
\and 
Th\"{u}ringer Landessternwarte Tautenburg, Sternwarte 5, D-07778 Tautenburg, Germany
\and 
ESA Estec, Keplerlaan 1,  2201 AZ Noordwijk, Netherlands
\and
Institute of Astronomy - K.U.Leuven, Celestijnenlaan 200D, B3001 Leuven, Belgium
}

\received{6 April 2010}
\accepted{Apr 2010}
\publonline{later}


\abstract{We present  the preliminary results of the study of an interesting target in the first CoRoT  exo-planet field (IRa1): CoRoT~102918586. Its light curve presents additional variability on the top of the eclipses, whose pattern suggests multi-frequency pulsations. The high accuracy
CoRoT light curve was analyzed by applying an iterative scheme, devised to disentangle the effect of eclipses from the oscillatory pattern.  In addition to the CoRoT photometry
we obtained low resolution spectroscopy  with the AAOmega multi-fiber facility at the Anglo Australian Observatory, which yielded a spectral classification as F0~V and allowed us to infer a value of the primary star effective temperature. The Fourier analysis of the residuals,  after subtraction of the binary light curve, gave 35 clear frequencies.
The highest amplitude frequency, of 1.22 c/d, is in the expected range for both $\gamma$ Dor and SPB pulsators, but the spectral classification favors the first hypothesis.  
Apart from a few multiples of the orbital period, most frequencies can be interpreted as   rotational splitting of the main frequency (an $\ell =2$ mode) and of its overtones. }
\maketitle
\begin{figure*}[t]
\includegraphics[width=146mm]{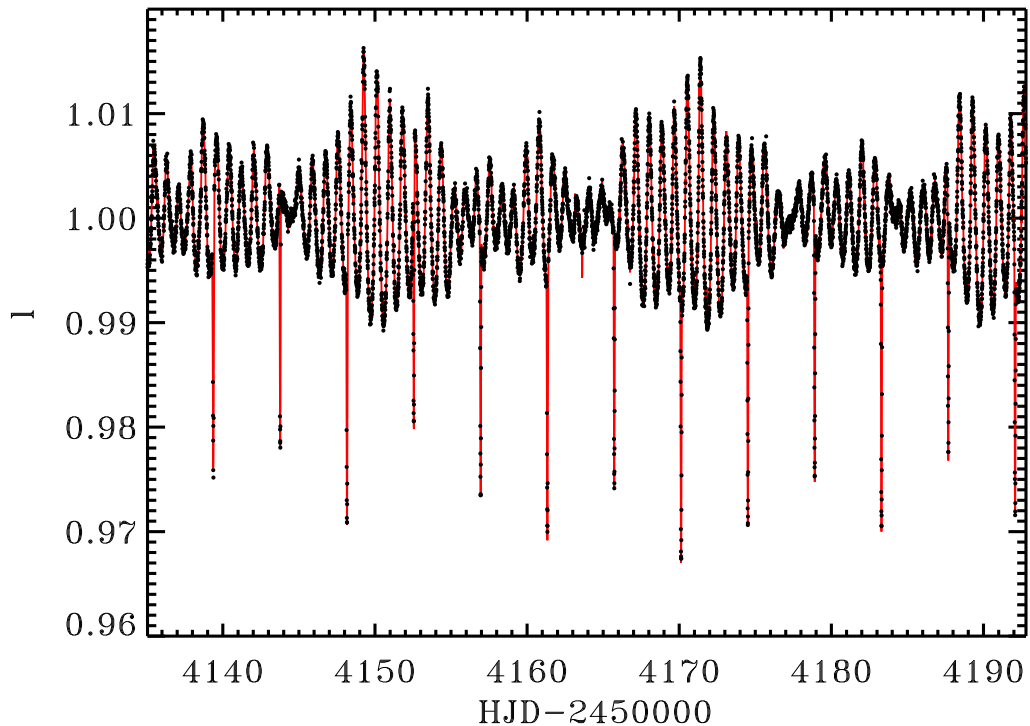}
\caption{The de-trended re-binned to 512$^{\mathrm s}$ (for better readability) light curve of CoRoT 102918586 and the final fit (full line) obtained by summing the binary model and the harmonic signal from the 35 detected frequencies.}
\label{label3}
\end{figure*}
\section{Pulsators in CoRoT eclipsing binaries}
The space mission CoRoT  \footnote{The CoRoT space mission was developed and is operated 
by the French space agency CNES, with participation of ESA's RSSD and Science Programs, 
Austria, Belgium, Brazil, Germany and Spain.} has already  provided high quality photometry and continuous monitoring of about 140000 stars.
Its photometric  accuracy ranges from a few to a few hundred parts per million (Auvergne et al. 2009), depending on star magnitude and on the observing channel (Asteroseismo or Exoplanet); the duration of the observing runs varies from  20 to 150 days (for short or long runs), with a typical duty cycle $>$ 90\%.  With these figures there is no surprise that most CoRoT targets show some kind of variability, often of  quasi-periodic character. In particular CoRoT has discovered and monitored hundreds of eclipsing binaries
 (EBs). Thanks to the Corot Variability Classifier (CVC, Debosscher et al. 2009), and a  further  detailed screening, we selected a group of EBs whose light curves present other periodic phenomena besides eclipses, possibly due to a pulsating component. Even though the analysis is made more complex by the superimposed phenomena, the benefit is the access to independent, and somewhat complementary, sour\-ces of information on the stellar fundamental parameters (double-lined spectroscopic binaries provide, for instance, a pure geometrical determination of masses and radii, which are most useful in asteroseismic studies).  
 
 CoRoT~102918586 is one of the most promising targets of our sample,
because of  its brightness (V$\sim 12.4$), which allows  spectroscopic follow-up, and of its  light curve shape clearly showing  modulated oscillations and narrow eclipses (Fig.~\ref{label3}).

\section{Method of analysis}\label{method}
All light curves of our sample need a preliminary treatment to remove possible instrumental trends and obvious outliers. When none of the two periodic phenomena (pulsation and orbit) can be treated as a perturbation, as   in the case  of CoRoT~102918586,  we apply the following iterative scheme:
\begin{itemize}
\item[A.] Fourier analysis  (with Period04,  Lenz \& Breger 2005), of  a time series obtained after removal of the eclipses. The frequency significance empirical criterion of Breger et al. (1993) is used. 
\item[B.]  Prewhitening of the de-trended full light curve (hereafter LC0) with the derived pulsation frequencies, providing a first-approximation binary light curve (LCB1)
\item[C.] Light curve solution based on LCB1 (or a conveniently resampled version) with WD-PHOEBE (Pr\v{s}a \& Zwitter  2005) and computation of the residuals from 
the full light cur\-ve LC0), yielding a new (and continuous) time series  to submit to Period04.
\item[D.] Second stage of prewhitening of LC0, providing LCB2, on which the PHOEBE solution can be refined. 
Steps B and C need to be repeated at least once, to refine the PHOEBE solution, but the loop can be repeated as many times as necessary.
\end{itemize}
\section{Observations of  CoRoT~102918586}
\subsection{CoRoT photometry}
CoRoT 102918586 was observed for about $57$ days (from $\mathrm{HJD} = 2454135.056$ to 2454192.499) during CoRoT's initial run in the exoplanet field (IRa1), with a time sampling of 512$^{\mathrm s}$ for the first 27 days and then of 32$^{\mathrm s}$ for the rest of the record. Little information was known about this target prior to the CoRoT mission, and its binary nature was revealed by the CoRoT Variability Classifier (Debosscher et al. 2009).

The light curve has a total of 87355 points, and is re\-cor\-ded in ``chromatic" light, i.e.~in three uncalibrated bands unevenly spanning the
wavelength interval 400-900 nm, obtained by the insertion of a bi-prism just before the CCD of the exo-planet channel. In this particular case, the three chromatic curves do not present any notable differences and we added them in order to obtain a single light curve with increased Signal to Noise Ratio (SNR). After the cleaning of obvious outliers, the removal of a slow decreasing trend due to some instrumental decay and homogenization of the sampling, we obtained the normalized curve shown in Fig.~\ref{label3}. The examination of the contaminants in the mask used to integrate the flux reveals that the brightest one is 3 magnitudes fainter than the target, and is, therefore, negligible. The light curve displays periodic variations in intensity presenting a beating pattern, with a maximum peak to peak amplitude of about 25 mmag that could be pulsations of one or both the components. Moreover, the eclipsing binary nature can clearly be seen thanks to deep regularly spaced minima. 
\subsection{Ground-based additional information}\label{sec:ground}
\begin{figure}
\includegraphics[width=83mm]{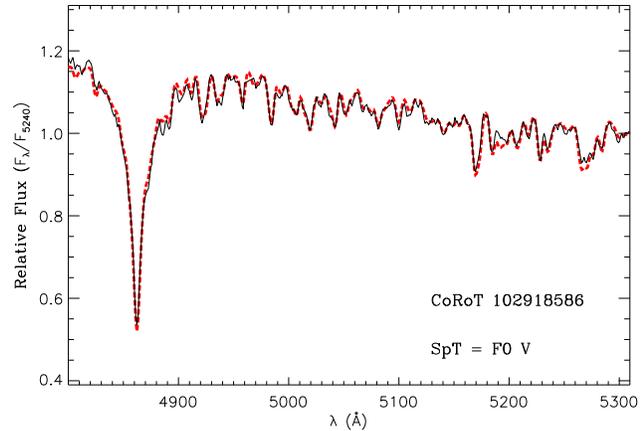}
\caption{ Low resolution spectrum (R$\sim$1300) of CoRoT 102918586 (black line), observed with the AAOmega multi-object facility at the Anglo-Australian Observatory. The best fitting F0~V template has been superimposed with a  broken line (T$_{\mathrm{eff}}=7300$ K, log g=4.0).}
\label{label1}
\end{figure}

The ExoDat database\footnote{The Exo-Dat database is operated at LAM-OAMP, Marseille, France, on behalf of the CoRoT/Exoplanet program} provides for this target the color information collected in table \ref{tab:EB7color}.
 \begin{table}[htb]
    \caption{Colour photometry for CoRoT 102918586}
     \label{tab:EB7color}
      \begin{center}
      	\begin{tabular}{lcccc}
         \hline
			&Harris B& Harris V&Sloan r'\\%
         \hline%
		Mag & $12.842\pm0.03$&$12.425\pm0.029$&$12.244\pm0.048$\\%
         \hline%
      	\end{tabular}
      \end{center}
	\end{table}	
In addition, we took advantage of the AAOmega multi-object facilities at the Anglo-Australian Observatory to acquire a first spectroscopic snap-shot of the CoRoT fields. Low resolution  spectroscopy (with $R \approx 1300$) of the CoRoT  targets in the Anticenter fields was carried out with the multi-fiber spectrograph during two observing runs, in January 2008 and December 2008 - January 2009. The spectrum obtained for CoRoT~102918586 is shown in (Fig. \ref{label1}) together with its best fit   F0~V template.
    
\section{The  eclipsing binary model}

Using the Phase Dispersion Minimization method (Stellingwerf, 1978) and refining the result with PHOEBE, we derived the orbital period  $P=8\fd78248$. Although a period of a half cannot be excluded from photometry alone, considering that: i)  there is no clear evidence of a secondary minimum corresponding to a $4\fd39$ period ; ii) after iteration (see section \ref{method}), a slight difference can be seen between minima (in the sense of larger dispersion in the secondary eclipse)   in accordance with the $8\fd78$ period,
 we decided to keep the value $P=8\fd78$. A final answer will come from the high resolution spectroscopy we plan to acquire for this target. 
 
\begin{figure}
\includegraphics[width=83mm]{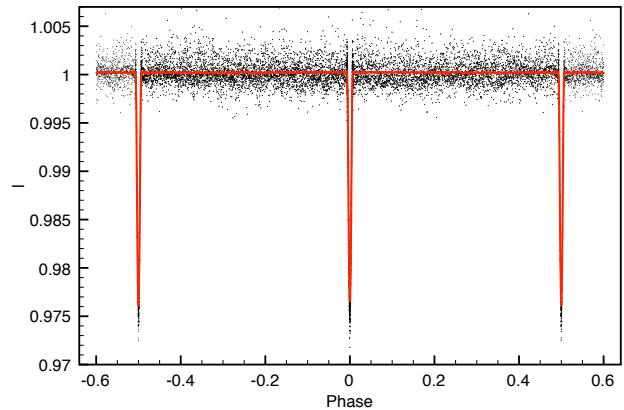}
\caption{The residuals pre-whitened with the final 35 frequencies with S/N $>$ 4, and yielding  the eclipsing binary light curve (LCB2). The original $\sim$75000 points are reduced in the LC solution to $\sim$21000, by resampling to 512$^\mathrm{s}$ bins outside eclipses.  The final solution is superimposed with a continuous line.}
\label{label2}
\end{figure}

 After pre-whitening with the fifteen frequencies de\-tec\-ted in step A of the iterative procedure (see section \ref{method}), we proceeded to the first LC solution. The original curve was re-sampled to 512$^{\mathrm s}$ bins in the out-of-eclipse intervals (to decrease computing time and the numerical weight of the large number of points in the flat sections). Step D (2nd pre-whitening) was then made. Degeneracy cannot be avoid while solving the inverse problem for eclipsing binaries, all the more so when there is only one input light curve, and the value of some parameters need to be assumed. Here we set the primary temperature $T_{\mathrm {eff,1}}$ at the one deduced from spectroscopy ($\pm$ 200 K) and assumed $m_2/m_1= 1$ (it cannot be derived from the LC solution of  detached systems). Albedo and gravity darkening were fixed at the theoretical values for radiative envelopes (1.0), and accordingly we used a logarithmic limb darkening law. The choice of the period implied a circular orbit (equidistant minima). The final PHOEBE solution is shown in Fig. \ref{label2} and the main parameters are given in Table~\ref{tab:EB7Param} (where $r$ is the fractional radius). The global fit (binary + pulsations) is shown in Fig.~\ref{label3}. All fits are obtained with the updated version of PHOEBE (Jan 2010) including model atmospheres and limb darkening tables in the CoRoT passband (Maceroni et al. 2009).
\begin{table}[htb]
\caption{Main parameters of the binary model. All errors are  formal errors derived from the least-squares fit.}
\label{tab:EB7Param}
\centering
\begin{small}
\begin{tabular*}{0.99\columnwidth}{@{\extracolsep{3pt}}l@{\extracolsep{3pt}}c@{\extracolsep{-5pt}}c@{\extracolsep{-5pt}}c}
\hline
					&                      		& System             &                      \\
                    & Primary            		&                    & Secondary            \\
\hline
$i$ (${}^\circ$)	&           		   & $ 86.053\pm0.001$     &                      \\
$T_\mathrm{eff}$ (K)& $7300^{a}$  &                 & $7350 \pm 10$      \\
$L/L_\mathrm{tot}$ & $0.5283 \pm 0.0001$             &                 & $0.4717 \pm 0.0001$  \\ 
$r$                &  $0.04081 \pm0.00001$             &                 & $0.0381\pm0.0001$              \\
\hline
$a)$ Fixed value
\end{tabular*}
\end{small}
\end{table}

\section{Pulsation analysis}

\begin{figure}
\includegraphics[width=83mm]{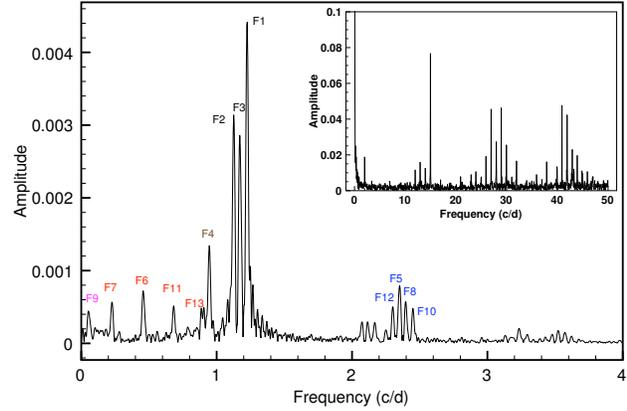}
\caption{ Fourier spectrum before pre-whitening, with indications of the first thirteen frequencies of Table~\ref{freq}, and the spectral window.}
\label{label4}
\end{figure}

The Fourier analysis of the residuals, after subtraction of LCB2 fit (Fig. \ref{label2}), yields 35 significant  frequencies. The first thirteen  are shown in Table \ref{freq} and Fig. \ref{label4}.
The frequency range, around 1 c/d, and the spectral classification suggest the presence of a  $\gamma$ Dor component. The eclipses being partial, we cannot tell which one is the pulsating component (though a decrease of dispersion at phase zero favors the hypothesis of a pulsating primary).

The straightforward interpretation of the frequency pattern is that we are observing the rotational splitting of an l=2 g-mode around $F_3$ and its overtones ($2 F_3$ and $3 F_3$), and a few multiples of the orbital frequency $F_{\mathrm{orb}}=0.1139$ c/d, which could be related to small systematic deviations of the binary light curve fit (Table \ref{freq_int}). The mean frequency splitting of about $0.05$ c/d corresponds approximatively to $0.5 F_{\mathrm{orb}}$. In a synchronized system - as a circular orbit suggests - $ F_{\mathrm{orb}}$ would also be the frequency of rotation. The splitting is indeed of the right order to be of rotational origin.
Note that $F_3/F_{\mathrm{orb}} \approx 10$, a similar ratio for tidally excited modes being already found in the SPB HD 177863 (Willems \& Aerts 2002) and HD 174884 (Maceroni et al. 2009)
\begin{table}
\caption{The first thirteen frequencies, ordered by amplitude (as detected by Period04  in the normalized light curve prewithened with the binary signal), their  corresponding phases  and S/N ratios. The numbers in brackets are the errors on the last digit. The S/N ratio was computed over an interval of 5 c/d. }
\label{freq}
\begin{tabular}[ht!]{lllll}
\hline
&F (c/d)& Ampl $\cdot 10 ^3$ &Phase (2$\pi$ rad) & S/N  \\ 
\hline
$F_1$&	1.22452 	(2)&	4.30116 (1)&	 0.1576 (3)	&24.7\\
$F_2$&	1.12666 	(3)&	3.05882 (1)&	 0.1223 (4)	&21.7\\
$F_3$&	1.17256 	(3)&	2.64945 (1)&	 0.6155 (5)	&24.2\\
$F_4$&	0.94653 	(4)&	1.43687 (1)&	 0.2809 (9)	&16.5\\
$F_5$&	2.3512 	(1)&	0.72474 (1)&	 0.101   (2)	&11.8\\
$F_6$&	0.4572 	(1)&	0.66486 (1)&	 0.432   (2)	&8.51\\
$F_7$&	0.2277 	(1)&	0.53671 (1)&	 0.286   (2)	&7.00\\
$F_8$&	2.39622 	(1)&	0.49724 (1)&	 0.903   (3)	&9.77\\
$F_9$&	0.0519 	(2)&	0.50943 (1)&	 0.251   (3)	&6.72\\
$F_{10}$&	2.4490 (2)& 0.41644 (1)&	 0.097   (3) 	&9.65\\
$F_{11}$&	0.6833 (2)& 0.44263 (1)&	 0.055   (3) 	&7.96\\
$F_{12}$&	2.3009 (2)& 0.42179 (1)&	 0.944   (3)	&9.33\\
$F_{13}$&	0.9118 (2)& 0.32531 (1)&	 0.901   (4)	&6.49\\
\hline 
\end{tabular}
\end{table}
\begin{table}
\caption{Frequency interpretation.}
\label{freq_int}
\begin{tabular}{lll|lll}
\hline
  &     F (c/d) &    Comment &  &  F (c/d) &Comment\\
\hline
$F_{25} $ &  1.074  &   q1, m=-2 &$F_{24} $ &  2.2499  &  q2, m=-2 \\
$ F_2 $ &  1.12666  &   q1,  m=-1 &$F_{12} $ &  2.3009  &  q2, m=-1 \\
$ F_3 $ &  1.17256  &   q1,  m=0 &$ F_5 $ &  2.3512  &  q2,  m=0 \\
$ F_1 $ &  1.22452  &   q1,  m=+1  &$ F_8$  &  2.39622  &  q2, m=+1 \\
$F_{20} $ &  1.2756  &   q1, m=+2 &$F_{10} $ &  2.4490  &  q2, m=+2 \\
&&&&\\
$F_{30}$   & 3.476  &   q3, m=-1 &$F_{7}$   &  0.2277   &  2 $F_{\mathrm{orb}}$ \\
$F_{23}$   & 3.5254  &   q3, m=0&$F_{6} $  &  0.4572   &  4 $F_{\mathrm{orb}}$ \\
$F_{27} $  & 3.5740  &   q3, m=+1 &$F_{11}$  &  0.6833   &  6 $F_{\mathrm{orb}}$ \\
$F_{35}$   & 3.622  &   q3, m=+2 &$F_{13}$  &  0.9118   &  8 $F_{\mathrm{orb}}$ \\

\hline
\end{tabular}
\end{table}

We expect to see $\ell=2$ modes excited in presence of a dynamical tide or an equilibrium tide, or both (e.g. Zahn, 2005). However, the first mechanism requires an eccentric orbit, whereas the second requires tidally deformed components. None of these conditions is fulfilled by the present binary model.
An alternative model could be built in the hypothesis of an orbital period half the assumed value and a particular orientation in space of an eccentric orbit, so that only one eclipse would occur in a system with similar components (SB2). A similar configuration is at the origin,
for instance,  of the light curve of HD 174884 (Maceroni et al. 2009) where twin components produce minima with a ratio of depths of about a hundred.

\section{Conclusions}
The preliminary results on CoRoT 102918586 are based on a single pass-band (though excellent) photometry and low resolution spectra, nevertheless the combination of pulsation and binarity yields already a rich harvest of information. Obviously we need high resolution and time resolved spectroscopy to definitely solve the remaining ambiguities of this interesting system. This target is included with high priority in our spectroscopic follow-up  programs.

\acknowledgements
 The research leading to these results has received funding from
 the Italian Space Agency (ASI) under  contract ASI/INAF I/015/07/00 in the frame of the ASI-ESS project,
from  the European Research Council under the European Community's Seventh 
Framework Programme (FP7/2007--2013)/ERC grant agreement n$^\circ$227224 (PROSPERITY),
as well as from the Research Council of K.U.Leuven grant agreement GOA/2008/04.
CM acknowledges   the European Helio and Asteroseismology Network (HELAS), a major 
internation\-al collaboration funded by the European Commission's Sixth Framework Programme, for supporting her contribution
to the conference.

\end{document}